%% file: DownWithBreitFrame.tex
\documentclass[10pt,aps,prd,amsmath,twocolumn,longbibliography]{revtex4-1}

\usepackage{amsmath}
\usepackage{graphicx}
\usepackage{amssymb,color}
\usepackage[utf8]{inputenc}
\usepackage[T1]{fontenc}
\usepackage{array}
\newcolumntype{P}[1]{>{\centering\arraybackslash}p{#1}}
\renewcommand{\bar}[1]{\overline{#1}}
\renewcommand{\bar}[1]{\overline{#1}}

\def\Dslash{\raise.15ex\hbox{/}\kern-.7em D}
\def\Pslash{\raise.15ex\hbox{/}\kern-.7em P}

\def\s{\sigma}

\newcommand{\la}{\langle}
\newcommand{\ra}{\rangle}
\newcommand{\bfr}{{\bf r}}

\newcommand{\ben}{\begin{displaymath}}
\newcommand{\een}{\end{displaymath}}
\newcommand{\be}{\begin{equation}}
\newcommand{\ee}{\end{equation}}
\newcommand{\bea}{\begin{eqnarray}}
\newcommand{\eea}{\end{eqnarray}}

\newcommand{\eq}[1]{Eq.~(\ref{#1})}

\newcommand{\bfp}{{\bf p}}
\newcommand{\bfP}{{\bf P}}\newcommand{\bfK}{{\bf K}}
\newcommand{\bfq}{{\bf q}}      \newcommand{\bfx}{{\bf x}}                                                       

\newcommand{\bfR}{{\bf R}}
      \newcommand{\bfD}{\mbox{\boldmath$\Delta$}}

\def\g{\gamma}\def\r{\rho}\def\L{\Lambda}\def\a{\alpha}\def\d{\delta}\def\l{\lambda}

\def\m{\mu}

\newcommand{\beqn}{\begin{equation}}
\newcommand{\eeqn}{\end{equation}}

\def\n{\nu}
 \def\D{\Delta}

\usepackage{textpos}

\begin{document}                                           

\begin{textblock*}{100mm}(.3\textwidth,-2cm)
NT@UW-25-8 
\end{textblock*}
 
\title{On the Impossibility of Obtaining Time-Independent, Three-Dimensional, Spherically-Symmetric  Densities of Confined Systems of Relativistically Moving Constituents 
}

\author{   Gerald A. Miller$^1$ }

\affiliation{$^1$ 
Department of Physics,
University of Washington, Seattle, WA 98195-1560, USA}
                                                                           
\date{\today}     
\begin{abstract}
The quantum mechanical definition of probability, the uncertainty principle and Poincar\'{e} invariance provide strong  basic restrictions on the ability to define spatial densities  associated with form factors describing the properties of confined systems of relativistically moving constituents. Despite this, many papers   ignore one or more of these restrictions. Here I show how to obtain {\it time-independent}, two-dimensional densities that are consistent with the stated restrictions. This is done using the light-front, infinite momentum frame formalism. Two-dimensional density interpretations of the axial-vector form factor and all three gravitational form factors are obtained. Additionally, an expression of a two-dimensional  mass density  related to the trace of the energy momentum tensor is obtained.  I also show that  all  known methods  for finding three-dimensional densities: using the Breit frame, Abel transformations, Wigner distributions and spherically-symmetric wave packets with vanishing spatial extent violate the basic restrictions in different manners. Furthermore,  the use of the latter    leads to densities that vanish almost everywhere in space as time increases from an initial value.  
\end{abstract}
\maketitle

\section{Introduction}
Due to recent exciting experimental developments, see the review ~\cite{Burkert:2023wzr}  and significant advances in lattice QCD calculations~\cite{Hackett:2023rif}, interest in gravitational form factors of hadrons and the related spatial densities has grown. Motivation for this interest continues to originate  from  the attempts to understand the proton mass puzzle~\cite{Ji:1994av,Ji:1995sv,Lorce:2017xzd,Hatta:2018sqd,Rodini:2020pis,Metz:2020vxd}
and the proton spin puzzle~\cite{Ashman:1987hv,Ji:1996ek,Leader:2013jra}

Matrix elements of the EMT between plane wave states
define gravitational form factors~\cite{Kobzarev:1962wt},
that provide information about both the quark-gluon decomposition
and the spatial distribution of energy, momentum, and angular momentum.
Those form factors are in principle measurable using    reactions
such as deeply virtual Compton
scattering~\cite{Ji:1996nm,Radyushkin:1997ki,Kriesten:2019jep}
at facilities such as Jefferson
Lab~\cite{Defurne:2015kxq,Jo:2015ema,Hattawy:2019rue}
and the upcoming Electron Ion Collider~\cite{Accardi:2012qut},
as well as in $\gamma\gamma^*\rightarrow\mathrm{hadrons}$
at Belle~\cite{Kumano:2017lhr}. 
Such studies complement the long-standing efforts, see {\it e.g.} the reviews~\cite{Arrington:2006zm,Perdrisat:2006hj,Carlson:2007sp}, to obtain spatial densities of electromagnetic properties.  

A large number of papers have been and are devoted to obtaining time-independent densities that obey spherical symmetry, meaning functions only of the spherical length coordinate $r$.    The principal statement  of the present  paper is that such densities can only be obtained for non-relativistic systems, those that  obey invariance under Galileo transformations. 
It is worthwhile to define non-relativistic in the present context:  the constituents move at speeds such that kinetic energies may be expressed as $p^2\over 2m$, and that the hadronic wave function is normalized as unity (instead of {\it e.g.} as twice the energy).

 For  some hadrons the constituents move at  speeds large enough so that Galileo transformations are not relevant.
 Nucleons, composed predominantly of $u$ and $d$ quarks with very small masses, provide the most important example.
  In such cases, one  must consider invariance under the Poincar\'{e} 
group instead of that of Galileo transformations. The  Poincar\'{e} 
group contains Lorentz transformations and rotations. A related problem is that form factors depend on
 Lorentz-invariant quantities, but the spatial coordinate $r$  is part of  a four-vector. A relativistic treatment of densities demands a consistent treatment of the time variable as well as the spatial variable. The net result is that it not possible to legitimately obtain a density that depends on $r$ and is independent of time.
 This paper is devoted   to explaining, in an indisputable fashion, the meaning of `legitimate' that appears in the  previous sentence.

I begin at the beginning with the postulate that  probability density is  determined by the absolute square of a wave function $|\Psi|^2=\Psi^*\Psi$. The key point is that   $\Psi^*$ is the complex conjugate of $\Psi$. The {\it same} wave function appears twice.  

 If a system obeys Gallileo invariance, the Hamiltonian can be   expressed as a sum of two separate terms, one involving internal coordinates and the other describing the motion of the center of mass~\cite{GY2003}. 
 In that case, the wave function factorizes into a  product of a wave function depending on internal, relative coordinate times a wave function depending on   the motion of the center of mass. 
 
 Next consider  a  Galileo-invariant    system   probed with a momentum transfer, $\D$. The action of the probe can be separated into terms depending on internal coordinates and the center of mass, and  the resulting transition amplitude is proportional to the three-dimensional Fourier transform of an appropriate density.
 This elegant result has been  ingrained in the physics consciousness of many physicists since the Nobel prize winning work of Hofstadter~\cite{PhysRev.102.851,Hofstadter:1957wk}. Initial applications were to the charge densities of nuclei, for which the treatment is valid~\cite{Miller:2009sg}. But problems with the interpretation of proton form factors as three-dimensional Fourier transforms of densities, arising from relativistic effects, were pointed out almost immediately by Yennie, L\'evy and Ravenhall~\cite{RevModPhys.29.144}. A modern perspective regarding the non-relativistic limit is given by  Freese~\cite{Freese:2025tqd}.

The quarks inside the proton and neutron move at speeds such that effects of special relativity are important~\cite{Miller:2009sg} so that Galileo invariance must be replaced by Poincar\'{e} invariance. 
The  Poincar\'{e} group comprises all geometric transformations that leave the quadratic form
$s^2(x_1,x_2)= c^2 (t_1^2-t_2^2)- (\bfr_1-\bfr_2)^2$ invariant~\cite{Stancu}. These transformations include rotations, Lorentz transformations or translations. The Lorentz group accounts for 3  pure Lorentz transformations and 3 rotations.
The group of space-time translations has 4 generations, so that the Poincar\'e group  consists of 10 generators. In the usual instant form,
translation generators  are  said to be dynamic because they contain interactions~\cite{RevModPhys.21.392}.   
The front form presents an advantage because only three translation generators contain interactions~\cite{RevModPhys.21.392,Kogut:1972di}. There is an additional kinematic translation generator.
This allows an important result to be obtained:   The  front form  ``is shown to have Galilean invariance with respect
to motions transverse to the direction" of the front~\cite{Susskind:1967rg}. So  there is two-dimensional Galileo invariance and    two-dimensional densities may legitimately  be obtained from two-dimensional Fourier transforms of form factors under two  conditions.

The first  condition involves the ability to define a position variable. For example,  the proton to neutron axial-vector current matrix element is given by
 \begin{widetext}
\bea _p\langle p',\l'|A_+^\mu(0)|p,\l\rangle_n=u_p(p',\l)[G_A(\D^2) \g^\m\g_5+{G_P(\D^2)\over 2M}\g_5\D^\m]u_n(p,\l)
,
\label{axial}\eea
\end{widetext}
where $\Delta^\m={p'}^\m-p^\m$, if  G parity 
is conserved.
 The related two-dimensional densities are discussed in Sec.~II. The difficulty is that the space-time position $(0)$ can be anywhere, everywhere and anywhen.  Plane-wave states, with  equal probability to exist anywhere in space appear in
 \eq{axial} so  that defining an origin is problematic. 
 The spread in momentum is zero, so according to the uncertainty principle,  the spread in coordinate space is infinite. One may instead use a wave packet that allows a   definite position in coordinate space to be defined.
 
 The second condition is that only kinematic  boosts,  those in the transverse direction to that of the light-front defined above, must be used.
 The light front coordinates are defined so that
$x^\pm = \frac{1}{\sqrt{2}}(x^0 \pm x^3)$ and $A\cdot B=A^+B^-+A^-B^+-{\bf A}_\perp\cdot{\bf B}_\perp$
 I choose $\sqrt{2}\D^+=\D^0+ \D^3$ to vanish, the Drell-Yan frame~\cite{PhysRevLett.24.181}.  This condition may be used for  space-like values of $\D$. 
  If $\D^+=0$ then $\D^2=-\D_\perp^2$, and one may define an appropriate wave packet~\cite{Burkardt:2000za,Freese:2021czn}. This means that the wave packet must be a superposition of nucleon states with the same  value of the plus-component of momentum for $\Psi$ and $\Psi^*$.

Here is an outline of the remainder of this paper. Sect.~II shows how to obtain time-independent two dimensional densities and  presents illustrative examples. 
Sect. ~III explains why the widely-used Breit-frame interpretation is severely flawed because it does  not satisfy the property that the density is the absolute square of a wave function~\cite{Miller:2018ybm} and because the uncertainty principle prevents a meaningful definition of position. 
Sect.~IV shows how the use of the Wigner distribution does not alleviate the problems associated with using the Breit frame because the fluctuations in both position and momentum are infinite.  Sect.~V explains that using the Abel transformation does not generally lead to a correct three-dimensional density~\cite{Freese:2021mzg} because of  length contraction effects for moving systems. Sect.~VI shows that the use of wave packets that are spherically symmetric  and of vanishing spatial extent leads to densities that vanish for times $t>r/c$ and  is inconsistent with standard calculations of form factors of relativistic systems. 
 The final Sect.~VII summarizes this paper.

\section{ Light Front Examples of Time-Independent Densities}

 I begin this section with some background information.  Much of this is known, see {\it e.g.}~\cite{Freese:2021czn},
  but the treatment of the time $(x^+)$ dependence is new, and I start by explaining the key point. 
  
  The sole advantage of using the Breit frame (Sect.~III)  is that the transferred energy vanishes, so that the four-momentum transfer becomes effectively a three-momentum transfer. The same feature is obtained by using the infinite momentum frame along with the Drell-Yan frame. To see this, note that  in light-front coordinates the Einstein energy-momentum relation for a free particle of four momentum $p^\m$ is given by 
  \bea
  p^-=\frac{\bfp_\perp^2+M^2}{2 p^+},
  \eea 
with $p^-$ playing the role of the energy. In the Drell-Yan frame only the perpendicular component of the target momentum is changed. Thus, by taking the limit $p^+\to\infty$ the transferred energy vanishes. This is one example of Feynman's statement that ``moving at large momentum $\cdots$ (interaction) times are dilated by the relativistic transformation so as $P$ rises things change more and more slowly" in Lecture 27 of Ref.~\cite{Feynman1972}.

 The limit must be taken with care, with $\bfp_\perp$ remaining finite. The necessary procedure will be  shown below. 

Many examples of two-dimensional densities that arise from using the infinite momentum frame along with the Drell-Yan frame have appeared, including ~\cite{Susskind:1968zz,PhysRevD.15.2617,Miller:2007uy,Pasquini:2007iz,Carlson:2007xd,HWANG2008345,PhysRevD.78.071502,PhysRevD.79.014507,TIATOR2009344,Miller:2009qu,PhysRevD.79.033003,Carlson:2009ovh,PhysRevD.81.013002,Dong:2010zza,Miller:2010nz,Strikman:2010pu,Venkat:2010by,Miller:2010tz,PhysRevC.88.065206,YAKHSHIEV2013375,Chakrabarti:2014dna,Carmignotto:2014rqa,Chakrabarti:2014dna,Mondal:2015uha,Chakrabarti:2016lod,Alarcon:2017asr,Mecholsky:2017mpc,Gramolin:2021gln}. The focus here is on obtaining time-independent results and presenting  two examples that have not yet been discussed. These are  the
  axial-vector density and a mass density  defined from the energy momentum tensor.

 Spatial densities in quantum field theories are given by expectation values
of local operators.
Given a local operator $\hat{\mathcal{O}}(x)$---such as the electromagnetic
current or the energy-momentum tensor---and a physically-realizable
single-hadron state $|\Psi\rangle$,
the relevant density is given by
$\langle\Psi|\hat{\mathcal{O}}(x)|\Psi\rangle$.

Form factors, $F_{\cal O}$,  are  associated with the local operator $\hat{\mathcal{O}}(x)$
through the matrix element between   one-particle plane wave states-
$\langle p',\lambda'|\hat{\mathcal{O}}(0)|p,\lambda\rangle$ for spin $1/2$ nucleons.
The connection between these two matrix elements is obtained by using the relation 
\bea 
\hat{\mathcal{O}}(x) = e^{i\hat{P}\cdot x}\hat{\mathcal{O}}(0)e^{-i\hat{P}\cdot x},
\eea
and evaluating matrix elements using a basis of  plane-wave states that are eigenfunctions of $\hat P$, the four-momentum operator.

The time $(x^+$) dependence  is kept   in addition to the usual spatial dependence.
Using light front coordinates  gives
\bea \hat{P}\cdot x =\hat{P}^-x^++ \hat{P}^+x^- - \hat{\mathbf{P}}_\perp\cdot\mathbf{x}_\perp
.\eea
I use   a change of variables
$P = \frac{1}{2}\big(p+p'\big)$ and $\Delta = p'-p$, with $\D^+=0$, twice use 
the completeness relation for single-nucleon states,  
\bea
  \sum_\lambda
  \int \frac{\mathrm{d}p^+\mathrm{d}^2\mathbf{p}_\perp}{2p^+(2\pi)^3}
  |p^+,\mathbf{p},\lambda\rangle\langle p^+,\mathbf{p}_\perp,\lambda|
  =
  1,
\eea
and then evaluate $\langle p',\lambda'|\hat{\mathcal{O}}(x)|p,\lambda\rangle$. This procedure
leads to
\begin{widetext}
\begin{multline}
  \label{eqn:232}
  \int \mathrm{d}x^- \,
  \langle\Psi|\hat{\mathcal{O}}(x^+,x^-,\mathbf{x}_\perp)|\Psi\rangle
  \\=
  \sum_{\lambda\lambda'}
  \int \frac{\mathrm{d}P^+\mathrm{d}^2\mathbf{P}_\perp}{2P^+(2\pi)^3}
  \int \frac{\mathrm{d}^2\boldsymbol{\Delta}_\perp}{(2\pi)^2} e^{i\frac{\bfP_\perp\cdot\boldmath{\Delta}_\perp x^+}{P^+}}
  \langle\Psi| P^+,\mathbf{p}'_\perp,\lambda'\rangle
  \frac{
    \langle P^+,\mathbf{p}'_\perp,\lambda' |
    \hat{\mathcal{O}}(0)
    | P^+,\mathbf{p}_\perp,\lambda \rangle
  }{2P^+}
  \langle P^+,\mathbf{p}_\perp,\lambda |\Psi\rangle
  e^{-i\boldsymbol{\Delta}_\perp\cdot\mathbf{x}_\perp}
  \,,
\end{multline}
\end{widetext}
  after integrating over all values of $x^-$,  necessary because boosts in the longitudinal direction are dynamic. This equation 
 is the general  space-time expression for an arbitrary nucleon state $|\Psi\rangle$.
For such a general state, the density associated with
$\hat{\mathcal{O}}(x)$ depends not only on the hadron's internal structure,
but also on its wave function spread.

A wave function that is completely localized at the origin is needed to obtain a density depending only on internal structure.
Localization may be  achieved by considering wave packets with
arbitrarily narrow width in the position representation,
though this width must be kept non-zero until after all
other calculations have been performed~\cite{Miller:2018ybm}.
A specific example---which is also a $P^+$ plane wave state,
and a state with definite light front helicity $\lambda$---is given by:
\begin{widetext}
\begin{align}
  \label{eqn:wf:lf}
  \langle p^+,\mathbf{p}_\perp,\lambda |\Psi\rangle
  =
  \sqrt{2\pi} (2\sigma) e^{-\sigma^2 \mathbf{p}_\perp^2}
  \,
  \sqrt{2p^+(2\pi) \delta_\epsilon(p^+-P^+) }
  \,
  \delta_{\lambda\Lambda}
  \,,
\end{align}
\end{widetext}
where $\delta_\epsilon(x)$ is an arbitrary representation of the
Dirac delta function
(such that $\delta_\epsilon(x)\rightarrow\delta(x)$
as $\epsilon\rightarrow0$).
The $\sigma\rightarrow0$ limit should not be taken until
  all momentum integrals have been performed.
 
Using the wave function of Eq.~(\ref{eqn:wf:lf}) in Eq.~(\ref{eqn:232}) gives:
\begin{widetext}
\begin{multline}
  \int \mathrm{d}x^- \,
  \langle\Psi|\hat{\mathcal{O}}(x^+,x^-,\mathbf{x}_\perp)|\Psi\rangle
 \\  =
  (2\pi)(2\sigma)^2
  \int \frac{\mathrm{d}^2\mathbf{P}_\perp}{(2\pi)^2} 
  e^{-2\sigma^2 \mathbf{P}_\perp^2}
  \int \frac{\mathrm{d}^2\boldsymbol{\Delta}_\perp}{(2\pi)^2}e^{i\frac{\bfP_\perp\cdot\boldmath{\Delta}_\perp x^+}{P^+}}
  \frac{
    \langle P^+,\mathbf{p}'_\perp,\Lambda |
    \hat{\mathcal{O}}(0)
    | P^+,\mathbf{p}_\perp,\Lambda \rangle
  }{2P^+}
  e^{-i\boldsymbol{\Delta}_\perp\cdot\mathbf{x}_\perp}
  e^{-\frac{\sigma^2}{2} \boldsymbol{\Delta}_\perp^2}
  \,,
\label{good0}
\end{multline}
\end{widetext}
  the most general   result for the $x^-$-integrated
density at $(x^+,\bfx_\perp)$. 

Note that  the $x^+$-dependence is contained in the factor $ e^{i\frac{\bfP_\perp\cdot\boldmath{\Delta}_\perp x^+}{P^+}}$, as expected. This and the Gaussian limit on the magnitude of $\bfP_\perp$ suggests  that the necessary frame to observe a time-independent density is the infinite momentum frame with $p^+,{p'}^+,P^+\to\infty$.

 Therefore I take the limit of the infinite momentum frame ($P^+\to\infty$)  to find:
 \begin{widetext}
\begin{multline}
  \int \mathrm{d}x^- \,
  \langle\Psi|\hat{\mathcal{O}}(x^+,x^-,\mathbf{x}_\perp)|\Psi\rangle_{\rm P^+\to\infty}
\equiv \r_{\cal O}(\bfx_\perp)\\  =
  (2\pi)(2\sigma)^2
  \int \frac{\mathrm{d}^2\mathbf{P}_\perp}{(2\pi)^2} 
  e^{-2\sigma^2 \mathbf{P}_\perp^2}
  \int \frac{\mathrm{d}^2\boldsymbol{\Delta}_\perp}{(2\pi)^2}
  \frac{
    \langle P^+,\mathbf{p}'_\perp,\Lambda |
    \hat{\mathcal{O}}(0)
    | P^+,\mathbf{p}_\perp,\Lambda \rangle
  }{2P^+}
  e^{-\frac{\sigma^2}{2} \boldsymbol{\Delta}_\perp^2}
  \,.
\label{good}
\end{multline}
\end{widetext}
The time-$(x^+$) dependence has disappeared!  The coordinate $x^+$ does not appear in the remainder of this paper.  
 
The ability to proceed  from \eq{good}  depends on the form
of the specific matrix element
$
    \langle P^+,\mathbf{p}'_\perp,\lambda' |
    \hat{\mathcal{O}}(0)
    | P^+,\mathbf{p}_\perp,\lambda \rangle
$.
If there are no such factors of  $\mathbf{P}_\perp$, doing  the
$\mathbf{P}_\perp$ integration, gives the result:
\begin{widetext}
\begin{align}
  \label{eqn:fourier:lf}
 \r_{\cal O}(\bfx_\perp)=
  \int \frac{\mathrm{d}^2\boldsymbol{\Delta}_\perp}{(2\pi)^2}
  \frac{
    \langle P^+,\mathbf{p}'_\perp,\Lambda |
    \hat{\mathcal{O}}(0)
    | P^+,\mathbf{p}_\perp,\Lambda \rangle
  }{2P^+}
  e^{-i\boldsymbol{\Delta}_\perp\cdot\mathbf{x}_\perp}
  e^{-\frac{\sigma^2}{2} \boldsymbol{\Delta}_\perp^2}
  \,.
\end{align}
\end{widetext}
This general density is a two-dimensional Fourier transform of a form factor 
\bea F_{\cal O}(\boldsymbol{\D})=\frac{\langle P^+,\mathbf{p}'_\perp,\Lambda |
    \hat{\mathcal{O}}(0)
    | P^+,\mathbf{p}_\perp,\Lambda \rangle
  }{2P^+}
  \eea
  If $F_{\cal O}$  
  depends only on the magnitude of $\boldsymbol{\Delta}_\perp$,
  then one may
  define a mean-squared transverse radius,
 \bea \la {\bfx_\perp^2}\ra_{\cal O}=\int d^2\bfx_\perp x_\perp^2  \r_{\cal O}(\bfx_\perp)\label{rad0}\\
 =-{1\over 4} {d\over d \boldsymbol{\D}_\perp^2}  F_{\cal O}(\boldsymbol{\D}_\perp^2).
 \label{rad}\eea
 The mean-squared transverse size  $ \la {\bfx_\perp^2}_{\cal O}\ra$ is a moment of a well-defined density, unlike the $\la r^2\ra$ of the Breit frame. The transverse squared radii related to a given form factor are two-thirds of the oft-used spherical radii.
An example of \eq{rad0} was derived in 1968~\cite{Susskind:1968zz}.

On the other hand, if the matrix element of $\cal O$ contains  factors of
$\mathbf{P}_\perp$, the resulting integral could contain a factor
$\sigma^{-1}$ or worse, causing the result to diverge in the
$\sigma\rightarrow0$ limit.
This essentially occurs because of the uncertainty principle:
an observable that depends on $\mathbf{P}_\perp$ cannot be well-defined
in the limit of the hadron's transverse position being arbitrarily well-known.

Two sets of examples of densities  that depend only on the internal structure of hadrons  (do not depend on the wave packet) are provided in the remainder of this section.

\subsection{Transverse Axial Current Densities} 

Given the preceding  background,  proceed to the case where $\hat{\mathcal {O}}(0)$ is
the axial vector current of \eq{axial}.   Lepage-Brodsky~\cite{Lepage:1980fj} spinors are used in the evaluations. Take $\m=+$, so that the second term on the right-hand-side vanishes and use the matrix element
$ {  \langle P^+,\mathbf{p}'_\perp,\Lambda |
   \g^+\g_5    | P^+,\mathbf{p}_\perp,\Lambda \rangle}
 = {2P^+}$ to obtain:
  \bea&
 \int dx^-\langle\Psi| A_+^+(0)(x^+,x^-,\mathbf{x}_\perp)|\Psi\rangle_{P^+\to\infty}\equiv   \r_A(x_\perp)\nonumber \\&=
  \int \frac{\mathrm{d}^2\boldsymbol{\Delta}_\perp}{(2\pi)^2}G_A(-\D_\perp^2)e^{-i \boldsymbol{\Delta}_\perp\cdot\bfx_\perp} \,.
  \label{ax} \eea
 The \eq{ax}  shows that the transverse axial charge density is simply the two-dimensional Fourier transform  of the axial form factor, $G_A$.

Obtaining the density associated with $G_P$ is more involved. 
First, concentrate on the second term of \eq{axial} and define the resulting density to be $\r_P(\bfx_\perp)$. Take the light front helicity state to be a superposition of positive and negative helicity states: $|\L\rangle=1/\sqrt{2}(|+\rangle+|-\rangle)$, meaning  polarization in the $x$ direction. The matrix element of $\g_5$ between $x$-polarized spinors is $-\D_x$. Then take $\m=x$. The result is that
\bea
\r_P(\bfx_\perp)=\nabla_x^2  \int \frac{\mathrm{d}^2\boldsymbol{\Delta}_\perp}{(2\pi)^2}G_P(-\boldsymbol{\Delta}_\perp^2)e^{-i \boldsymbol{\D}_\perp\cdot\bfx_\perp}\nonumber\\
=\tilde G_P''(x_\perp)\cos^2\phi+ \frac{\tilde G_P'(x_\perp)}{x_\perp}\sin^2\phi, 
  \eea
  where $\tilde G_P(x_\perp)$ is the two-dimensional Fourier transform of $G_P$, and $\phi$ is the angle between $\bfx_\perp$ and the direction of polarization.

\begin{figure}[ht]
  \centering
  \includegraphics[width=0.4\textwidth]{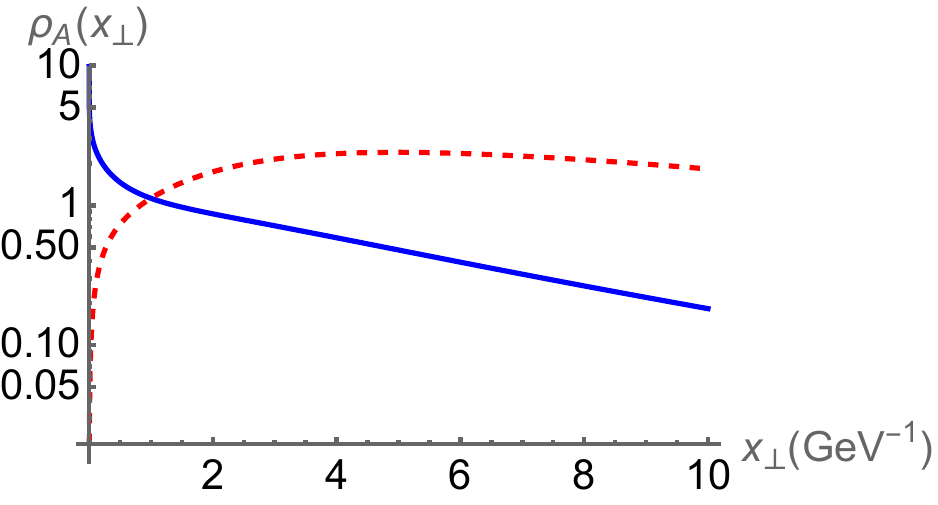}
  \caption{Solid- $\tilde\r_A(b)$, Dashed-$b \tilde\r_A(b)$ } 
\label{rhoA}  \end{figure}

\begin{figure}[ht]
  \centering
  \includegraphics[width=0.4\textwidth]{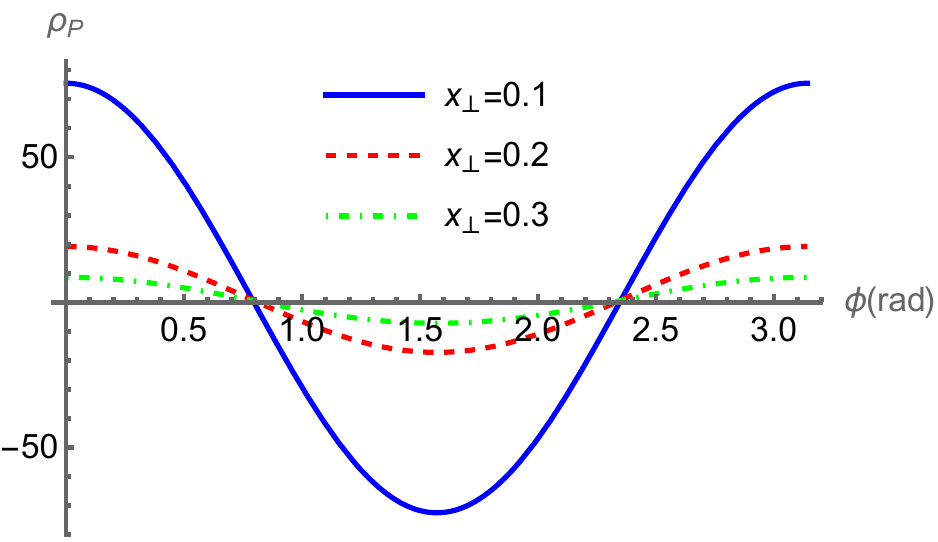}
  \caption{$\r_P(\phi)$ for $b=0.1,0.2,0.3$ GeV$^{-1}$.}
\label{rhoP}  \end{figure}

What do $\r_A$ and $\r_P$ look like?  The   dipole form of Ref.~\cite{MINERvA:2023avz}  is used  to provide an illustration. Thus
\bea 
G_A(\D^2)=\frac{G_A(0)}{(1-\D^2/M_A^2)^2},
\eea
with $G_A(0)=1.2723$ and  $M_A=1.014 $ GeV.
Then
\bea \r_A(x_\perp)=G_A(0)M_A^2\frac{M_A x_\perp\, K_1(M_A x_\perp)}{4 \pi },
\eea
where $K_1$ is a  modified Bessel function.
A simple form of $G_P(Q^2)$ is obtained if this induced pseudoscalar term has a pion pole. Then  the Goldberger-Treiman relation is generalized   to non-zero values of momentum transfer~\cite{Thomas:2001kw,Alberg:2012wr}:
\bea
G_P(\D^2)= \frac{4M^2 G_A(\D^2)}{-\D^2+m_\pi^2}.
\eea
Then taking the two-dimensional Fourier transform leads to the result:
\begin{multline}
\tilde G_P(x_\perp)=\\G_A(0)
\frac{M^2 \left(2 K_0\left(m_\pi x_\perp \right)+M_A x_\perp (m_\pi^2/M_A^2-2) K_1(M_A x_\perp)\right)}
{4 \pi    (1-m_\pi^2/M_A^2)^2}. 
\end{multline} 
The transverse axial charge density $\tilde G_A$ and  $x_\perp\tilde G_A$ are shown in Fig.~\ref{rhoA}.
The induced pseudoscalar density, $\r_P(\bfx_\perp)$, is shown in Fig.~\ref{rhoP} to have   a very strong spatial anisotropy.

\subsection{Densities of gravitational form factors}

The proton gravitational form factors $A(t), J(t)$, and $D(t)$  are defined~\cite{Kobzarev:1962wt} as
\begin{widetext}
\begin{align}
  \langle p',\lambda |
  T^{\mu\nu}(0)
  |p,\lambda\rangle
  =
  \bar{u}(p',\lambda) \left\{
    \frac{P^\mu P^\nu}{M} A(t)
    +
    \frac{\Delta^\mu\Delta^\nu-\Delta^2g^{\mu\nu}}{4M} D(t)
    +
    \frac{iP^{\{\mu}\sigma^{\nu\}\rho}\Delta_\rho}{2M} J(t)
    \right\}
  u(p,\lambda)
  \,,
\label{Tmunu}
\end{align}
\end{widetext}
where $P=(p+p')/2, \Delta=p'-p$, $t=\D^2$, $\s^{\m\n}=i/2[\g^\m,\g^\n]$,
and the curly brackets $\{\}$ signify symmetrization over the indices
between them (without a factor $1/2$). $T^{\m\n}$ is the renormalization-scale-independent symmetric energy momentum tensor of QCD.

I wish to obtain spatial  densities related to each of the form factors appearing in \eq{Tmunu}. One may isolate $A(t)$ by considering the matrix element of $T^{++}(0)$. The result is 
\bea
  \langle p',\lambda |
  T^{++}(0)
  |p,\lambda\rangle
  =2P^+ P^+ A(t)
  .\eea
Using this matrix element for $\hat{\cal O}(0)$ in \eq{good} and taking $\s\to0$ leads to the density:
\begin{widetext}
\begin{align}
\r_A(\bfx_\perp)\equiv {1\over P^+} \int \mathrm{d}x^- \,
  \langle\Psi|\hat T^{++}(x^-,\mathbf{x}_\perp)|\Psi\rangle
  =
  \int \frac{\mathrm{d}^2\boldsymbol{\Delta}_\perp}{(2\pi)^2}
   A(\D^2)
  e^{-i\boldsymbol{\Delta}_\perp\cdot\mathbf{x}_\perp}.
\end{align}
\end{widetext}
The term $\r_A(\bfx_\perp)$ has been denoted the momentum density~\cite{PhysRevD.78.071502}. Its integral is unity.

The spatial densities associated with $D(t)$ are those of the pure stress tensor. Keeping the second term on the right-hand side of \eq{Tmunu}, defining its contribution to be that of the matrix element of $T_D^{\m\n}(0)$ and taking the $i,j=1,2$ components allows one to use \eq{good} to define spatial densities~\cite{Freese:2021czn}. (Using any term that involves a minus-component would lead to divergent terms in the integral over ${\bf P}_\perp $ that appears in \eq{good}.) The result is 
\begin{widetext}
\begin{align}
\r_D^{ij}(\bfx_\perp)\equiv {  P^+} \int \mathrm{d}x^- \,
  \langle\Psi|\hat T_D^{ij}( x^-,\mathbf{x}_\perp)|\Psi\rangle
  ={1\over 4}
  \int \frac{\mathrm{d}^2\boldsymbol{\Delta}_\perp}{(2\pi)^2}
  (\D^i\D^j+\bfD_\perp^2 \d^{ij})D(\D^2)
  e^{-i\boldsymbol{\Delta}_\perp\cdot\mathbf{x}_\perp}\,.
\end{align}
\end{widetext}
 
Next, turn to the transverse density  associated with the proton's angular momentum, $J(t)$.  Keeping the third term on the right-hand side of \eq{Tmunu}, defining its contribution to be that of the matrix element of $T_J^{\m\n}(0)$ and taking the component $T_J^{+j}$  with $j=1,2$  allows one to use \eq{good} to define spatial densities~\cite{Freese:2021czn}.
This is because  
\bea
 \bar{u}(p',\lambda)  
       \frac{iP^+\sigma^{j\rho}\Delta_\rho}{2M}  
  u(p,\lambda)
  \,
  = i (\hat \lambda \times \bfD_\perp)^jP^+, 
\eea
where $\hat \lambda$ indicates the direction of the light-front helicity in the light-front direction. 
This result is obtained using the Gordon identity and Lepage-Brodsky spinors. Then using $T_D^{+i}(0)$ in \eq{good} leads to the result:
\bea&
\r^j_J(\bfx_\perp,\l)\equiv \int dx^-\langle \Psi|T_J^{+j}(x^+=0,x-,\bfx_\perp)|\Psi\rangle\nonumber\\&={1\over2}(\hat \l\times \nabla_\perp)^j \tilde J(|\bfx_\perp|),
\eea
where $\tilde J(|\bfx_\perp|)$ is the two-dimensional Fourier transform of $J(\D^2)$.
 
The three densities have been determined by using components of $T^{\m\n}$ that allow the integral over $\bfP_\perp$ to be performed with a divergence-free result.

There is one more combination to consider here, the trace of the energy momentum tensor, $T^\m_\m$. Take 
${\mathcal{O}}(0)$ to be $T_\mu^\mu(0)$ and use it in \eq{good} along with~\cite{Ji:2021mtz}  
\begin{widetext}
\bea
&\bar u(p',\l)T^\m_\m(0)u(p,\l)=2M G_s(\Delta^2 )
    =  2(M^2-\D^2/4)A(\Delta^2)   + J(\Delta^2){\Delta^2} - {3\D^2\over 4} D (\Delta^2) ,
\eea
\end{widetext}
to find
\begin{eqnarray}
  &{ m(x_\perp) }\equiv  {P^+\over M}  \int dx^-\langle \Psi|T^\mu_\mu(x^-,{\bf x}_\perp)|\Psi\rangle\nonumber\\&= \int \frac{d^2 \boldsymbol{\Delta}_{\perp}}{(2\pi)^2}\,G_s(\D^2)\, e^{-i\boldsymbol{\Delta}_{\perp}\cdot \boldsymbol{x}_{\perp}}  
\end{eqnarray}
The key feature in obtaining this result is that the matrix element of $T^\m_\m$ is independent of $\bfP_\perp$.
Note that $\int d^2x_\perp \, m(x_\perp)=2M^2$, so that $ m(x_\perp)$ may be thought of as a mass (squared) density.

 \begin{figure}[ht]
  \centering
  \includegraphics[width=0.45\textwidth]{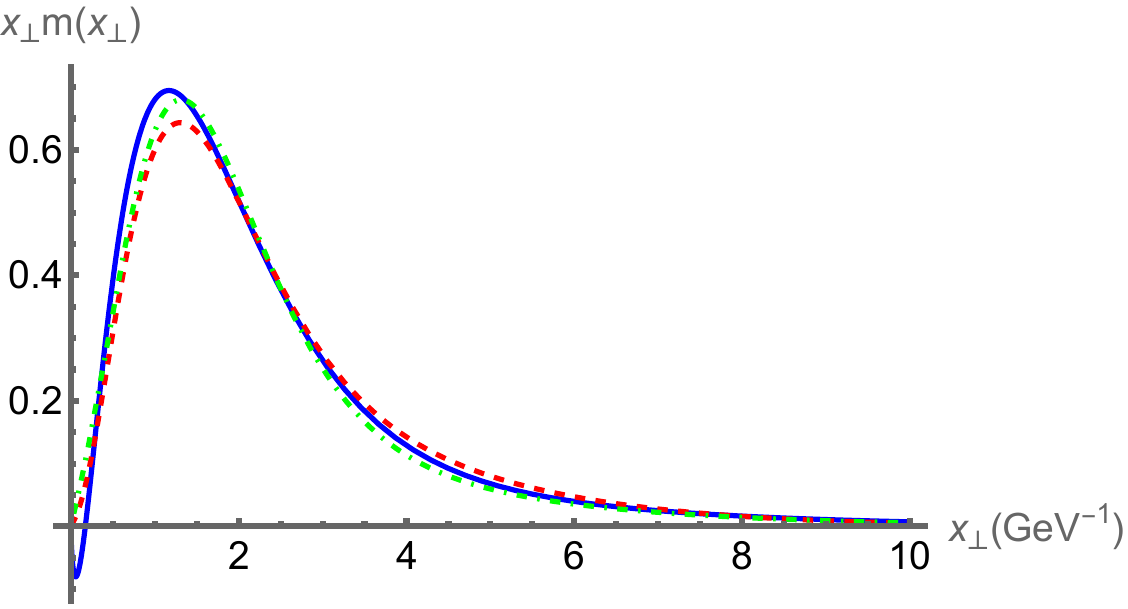}
  \caption{Mass (squared) distribution $m(b)$ Solid- dipole, Dashed-tripole, dot-dashed 4-pole .  }
\label{mofb}  \end{figure}

The resulting values of $x_\perp m(x_\perp)$, Fig.~\ref{mofb}, are 
 obtained  by using the lattice QCD  n-pole  form factors, with $n=2,3,4$, of  \cite{Hackett:2023rif} and the supplementary material.
The short distance values are very sensitive to the high momentum form factors, and the use of the dipole   leads to negative values of $ m(x_\perp) $ for small values of $x_\perp$. 
The dominant  change between using $n=2$ and $n=3$ comes from the change in $D(\D^2)$.  Calculations of perturbative QCD favor using a dipole for $A$ and $J$ and a tripole for $D$~\cite{Tanaka:2018wea,Tong:2021ctu}
.\\

The root-mean-square transverse radii (square root of the mean-square transverse radii of \eq{rad}) are
$0.60,\,0.54$ and 0.52 fm for $n=2,3,4$ respectively. These   seem to be small, but recall that these are smaller by a factor of $\sqrt{2/3}$  than conventional expressions of radii, due to using two  dimensions instead of three. 

The remainder of this paper is devoted to showing that obtaining three-dimensional densities related to form factors of confined systems of relativistically moving constituents   is not justified. Recent unfortunate examples include Refs.~\cite{Hackett:2023rif,Meziani:2025dwu}.
\section{What's Wrong with the Breit frame interpretation of densities}

The standard formulation~\cite{Halzen:1984mc,Sachs:1962zzc} of the Breit frame interpretation is discussed first. The Breit frame is defined for space-like probes by having the initial and final proton momentum being equal in magnitude and opposite in direction, so that no energy is transferred, $q^0=0$. This enables $q^2$ to be expressed as $q^2=-\bfq^2$, and taking a   three-dimensional  Fourier transform is possible. The matrix element of the 0'th component of the electromagnetic current is proportional to the Sachs electric form factor  $G_E$, {if the helicities of the initial and final states are opposite}.

Then one makes a non-relativistic treatment (see \cite{Halzen:1984mc}, Chapt. 8.1) to take a three-dimensional Fourier transform of $G_E(\bfq)$ to obtain a function of the three-dimensional distance coordinate. 
Using Galileo invariance allows the separation of relative and internal variables  so that this procedure is valid
\cite{Miller:2010nz}. However, this separation is not valid for a system in which the component particles move relativistically.
 An attempted derivation of this procedure  for relativistic systems was presented in~\cite{Sachs:1962zzc}, but this  was fatally flawed because an infinite term was neglected~\cite{Miller:2018ybm}.

  \subsection{Frame-dependent internal wave function}
A specific simple example of the frame-dependence of an internal wave function  is obtained by considering a hadronic  bound-state system of two light spin-0  particles.
 Then one may compute a form factor using a
Bethe-Salpeter wave function
$\Phi(k;P)$, where $k$ is the relative four-momentum and $K$ is the
total four-momentum. Such  can be written in a compact form if the
interaction kernel is obtained by summing  a set of
Feynman graphs~\cite{Nakanishi:1969ph,Nakanishi:1988hp}.
Then
the Nakanishi integral representation:
\cite{Nakanishi:1963zz} provides an explicit expression of the wave function:
\begin{widetext}
\begin{eqnarray}\label{bsint}
\Phi(k;K)&=&-{i\over \sqrt{4\pi}}\int_{-1}^1dz\int_0^{\infty}d\gamma
\frac{g(\gamma,z)}{\left[\gamma+m^2
-\frac{1}{4}M^2-k^2-K\cdot k\; z-i\varepsilon\right]^3},
\end{eqnarray}
\end{widetext}
where $m$ is the mass of each constituent scalar particle and $M$ is the
hadronic mass.
The weight function 
$g(\gamma,z)$ 
itself is not singular, but 
 the singularities of the BS amplitude are  reproduced by using \eq{bsint}
 \cite{Carbonell:2008tz}.

Consider the simplest case that $g(\g,z)=1$ to illustrate the key point. Then integration reveals that
\bea
 \Phi(k;K)&=&-{i\over \sqrt{4\pi}}\frac{1}{(m^2-M^2/4-k^2-i \epsilon)^2-(K\cdot k)^2},\nonumber\\
 \label{ex}
 \eea
 which can be shown to be a product of propagators~\cite{Miller:2009fc}. Note that this expression cannot be factored into a product of a function of $K$ with another function of the relative momentum, $k$, so that one obtains  different $k$-dependence for every value of $K$.

This explicit dependence on $K$ 
dramatically influences our understanding of
form factors because  
  the initial and final hadrons  have different momentum, $K$ and $K+q$,
and  therefore different   relative wave functions. 
The presence of different wave functions of  the initial and final
nucleons invalidates  a  
  probability or density  interpretation. In the Breit frame the nucleon initial and final momentum are opposite, so that the associated wave functions of the two states are different.

 That   the initial momenta   is  different from the final value is a severe problem, but there is something even worse-the violation of the uncertainty principle!
A plane wave state of definite momentum  has a constant probability density to be anywhere and everywhere  in space. Defining a position requires that one use a wave packet that is a superposition of momentum eigenstates. 
Since  $\bfP=({\bf K}+{\bf K}')/2=0$ appears with vanishing probability  in such superpositions, the Breit frame is not relevant.

\section{What's Wrong With Using the Wigner Distribution}

In Ref.~\cite{Lorce:2018egm} the authors present the three-dimensional spatial distributions defined
in the Breit frame where the system is on {\it average} at rest. The aim of this Section~IV  is to show that this criteria is not sufficient to define spatial positions of a system that is at rest because  the fluctuations in the values of position and momentum are infinite.
   
I begin by reviewing the procedure of Ref.~\cite{Lorce:2018egm}  for distributions in the instant form, using its  notation. The phase space density for an average position $R$ and average momentum $P$   is given by
\begin{equation}\label{RPnoncov}
\rho_{R,P}=\int\frac{d^3\bfD}{(2\pi)^3\,2P^0}\,e^{-i\Delta\cdot R}\,|{P-\tfrac{\Delta}{2}}\rangle \langle{P+\tfrac{\Delta}{2}}|\,,
\end{equation}
where the initial and final target energies are given by   on-mass-shell values. The aim of using the Wigner distribution
is to   ``localize the system in both position and momentum space".  This is impossible if one accepts the uncertainty principle.

Matrix elements of position-dependent operators $\hat {\cal O}$ using the Wigner distribution are  given by
\begin{equation}
\langle \hat{\cal O}(X)\rangle_{R,P}={\rm Tr}[\hat{\cal O}(X)\rho_{R,P}]\,,
\label{trace}\end{equation}
with  translation invariance implying that
\begin{equation}
\langle O(X)\rangle_{R,P}=\langle O(x)\rangle_{0,P}\,,
\end{equation}
where $x=X-R$ is the relative average position.

Specifying to  the energy momentum tensor, setting the origin at the average position of the system $\bfR={\bf 0}$ and denoting $x^\mu=(0,\bfr)$, the static EMT encoding the distribution of energy and momentum inside the system with canonical polarization $\bf s$ and average momentum $\bfP$, is {\it defined} to be  the following Fourier transform
\begin{equation}\label{TFT}
\mathcal T^{\mu\nu}(\bfr;\bfP)=  
\int\frac{d^3\boldsymbol{\D}}{(2\pi)^3}\,e^{-i\boldsymbol{\D}\cdot\bfr}\,\frac{\langle{p',{\bf s}} |T^{\mu\nu}(0)|{p,{\bf s}\rangle}}{2P^0}.
\end{equation} 
Thus, the resulting densities are simply a definition. 
Note that $x^0=0$ because the positions $\bf X$ and $\bf R$ are chosen to be at the same time $X^0=R^0$.

I  next show that although the average of the  momentum is  indeed  0, the fluctuations are infinite. This is done  by using 
 \eq{RPnoncov} and  the trace, \eq{trace}, to compute the average momentum and the average squared momentum.
Then the average momentum is given by
\begin{widetext}
\bea 
&\la \bfK\ra =\int d^3K  \bfK\int\frac{d^3\boldsymbol{\D}}{(2\pi)^3\,2P^0}\,e^{-i\Delta\cdot R}\,\la K |P-\D/2\ra\la P+\D/2|K\ra 
= \int d^3K {\bfK \over E(K)}=0,
\eea
\end{widetext}
where $E(K)=\sqrt{K^2+M^2}$.
One indeed obtains the result that the average value of the momentum vanishes, but only as a consequence of parity. So it is no surprise that  using the same formalism yields
 \bea 
&\la \bfK^2\ra = \int d^3K {\bfK^2 \over E(K)}=\infty.
\eea
The fluctuations in the momentum are infinite. The statement that the average momentum vanishes is true, but that is not sufficient to provide a meaningful state at rest.

Next I turn to coordinate space with the average position obtained as:
\bea &\la \bfr\ra=\int d^3r \bfr  \frac{d^3\boldsymbol{\D}}{(2\pi)^3\,2P^0}\la\bfr|P-\D/2\ra\la P+\D/2|\bfr\ra e^{i\boldsymbol{\D}\cdot\bfr}. \nonumber\\
\eea
The necessary overlap matrix element  is given by
\bea
\la\bfr|P-\D/2\ra=e^{i (\bfP-\boldsymbol{\D}/2)\cdot\bfr}
,\eea
obtained by using the normalization of $|x\ra$ of Eq (2) of~\cite{Lorce:2018egm},
\bea
|x\ra=\int {d^3p\over 2E(p) (2\pi)^3}e^{-i \bfp\cdot\bfr}|p\ra
\label{xnorm}\eea
 with $|x\ra$ having  dimension of inverse length squared.
Then 
\bea
&\la \bfr\ra= 
\int d^3 r \bfr \int {d^3\D\over (2\pi)^3}  {1\over E(\bfP+\boldsymbol{\D}/2) +E(\bfP-\boldsymbol{\D}/2)}=0,
\eea
where again spatial inversion is used to obtain the vanishing. The promise Ref.~\cite{Lorce:2018egm}  of a vanishing quantity is kept.

Next consider the fluctuations in position. Using the previous manipulations one finds that 
\bea& \la \bfr^2\ra\nonumber\\&=\int d^3 r \bfr^2 \int {d^3\D\over (2\pi)^3}  {1\over E(\bfP+\boldsymbol{\D}/2) +E(\bfP-\boldsymbol{\D}/2)}=\infty.\nonumber\\
\label{fr}\eea
 The fluctuations in position are  also infinite in any reference frame. 

Another noteworthy feature of the preceding equation is that the dimensions on the right-hand-side are of length cubed.
This peculiarity arises from the definition of the position eigenstate, \eq{xnorm}. Replacing the factor $1\over 2 E(p)$ by
$1\over\sqrt{2 E(p})$ would lead to expressions with the proper dimensions and remove the energy denominator from \eq{fr}. The fluctuations in position are still infinite.
  
Although the average position and momentum vanishes, the fluctuations in both quantities are infinite. Thus, there is  no state that is both at fixed position and is at rest.

 \section{The Abel transformation does not yield a proper three-dimensional density}
 Several authors including~\cite{Panteleeva:2021iip,Kim:2021jjf,Kim:2022syw,Kim:2021kum,Choudhary:2022den,Choudhary:2023ihs} have used the Abel transformation in an attempt to obtain three-dimensional spherically symmetric densities.
 Ref.~\cite{Freese:2021mzg} convincingly showed that  ``the application of the inverse Abel transform to the light front densities does not produce a physically
meaningful result."  
I  
  include a few remarks in the interest of presenting a complete description of why three-dimensional, spherically symmetric densities of confined systems of relativistically moving constituents do not exist. 

The Abel transformation is an integral transform that is used in analyzing spherically symmetric functions. Suppose there is a system that has a density, $\r$,  that is spherically symmetric. Then one may integrate along a chosen direction, $z$, to obtain a function of the coordinates transverse to $z$:
\bea\r(b)=\int_{-\infty}^{\infty} dz \r(r)=2 \int_b^\infty dr {\r(r)\over \sqrt{b^2-r^2}},\eea
in which $r=\sqrt{b^2+z^2}$. 
The first term is familiar from the Glauber (eikonal) theory of scattering. It gives the thickness of matter traversed by a beam moving through matter on a straight line trajectory with an impact parameter $b$, and indeed $\r(b)$ is denoted as the thickness function in many applications. 

 The immediate problem with applying an Abel transformation to understand  confined systems of relativistically moving constituents, is that such  a moving system  cannot have a spherically symmetric density because of the effects of Lorentz contraction. 
 
 Note also that there  are significant EIC-related efforts to determine the three-dimensional structure of the nucleon. It is universally known that the three directions are the longitudinal momentum, $x=k^+/p^+$, and the position of quarks and gluons 
in the transverse spatial plane, see {\it e.g.} page 99 of ~\cite{AbdulKhalek:2021gbh} and also {\it e.g.}~\cite{Li:2022hyf}.
The longitudinal and transverse directions are described using different variables, so that there can be no spherical symmetry. More generally, one may say that
the non-spherical nature of the nucleon structure is due to the requirements of Poincar\'{e} invariance

It is possible to  obtain wave functions and densities in three-spatial directions by Fourier transforming the longitudinal momentum to the Miller-Brodsky~\cite{PhysRevC.102.022201} longitudinal spatial coordinate,  $\tilde z= p^+x^-$, conjugate to $x$. Some examples, show that a two-particle wave function can be a product of a function of $\bfx_\perp$ with another different function of $\tilde z$. The lack of   spherical symmetry renders the Abel transform inapplicable.
 
 \section{Using spherical wave packets with vanishing spatial extent does not help} 
The work of Ref.~\cite{PhysRevLett.129.012001} states ``that the matrix element of a local operator between hadronic states can be used to unambiguously define the associated spatial density.".  The relation between the charge density of a spinless particle and the electric form factor is used as an example in that paper.

The analysis of Sect.~II, based on work done since 1968, shows  already how to obtain spatial densities for systems  having relativistic dynamics. Nevertheless, it is worthwhile to examine the interesting claim of the previous paragraph by examining the details of Ref.~\cite{PhysRevLett.129.012001}.

The authors of Ref.~\cite{PhysRevLett.129.012001}, use a frame in which the average momentum of the initial and final hadrons is zero (ZAMF), meaning $\bfP=0$. Such a frame is, of course, a Breit frame. This is because the initial and final nucleons must have the same energy. These authors improve on the standard use of the Breit frame
 by using wave packets. However, they state,  ``we define
the charge density distribution in the ZAMF by employing
spherically symmetric wave packets". Thus, their resulting densities are simply a definition.   Furthermore, such densities explicitly depend on the chosen wave packet. There is no reason why spherically symmetric wave packets must be used.  See, for example, 
 \eq{eqn:wf:lf}.

\subsection{The vanishing spherically symmetry density}

But there is a more fundamental problem with the approach of  Ref.~\cite{PhysRevLett.129.012001}, starting with the second  equation of that paper, reproduced here:
\bea \la p'|\hat\r(\bfr,0)|p\ra=e^{-i(\bfp'-\bfp)\cdot\bfr} (E+E') F(q^2)
.
\label{ulf}
\eea
In the ZAMF $E'=E=\sqrt{\bfq^2/4+M^2}$ and $\bfq^2=-(\bfp'- \bfp)^2$.
One may then Fourier transform \eq{ulf} to find
\bea
F(q^2)=\frac{1}{2E(2\pi)^3} \int d^3re^{i(\bfp'-\bfp)\cdot\bfr} \la p'|\hat\r(\bfr)|p\ra.
\label{ulf1}\eea
This expression looks suspect because the left-hand-side is a scalar, independent of reference frame, but the right-hand-side depends on $\bfq=\bfp'-\bfp$, a three-vector and $E$, the time-component of a four-vector.

\begin{widetext}
\begin{figure}[ht]
\centering
\includegraphics[width=1\textwidth]{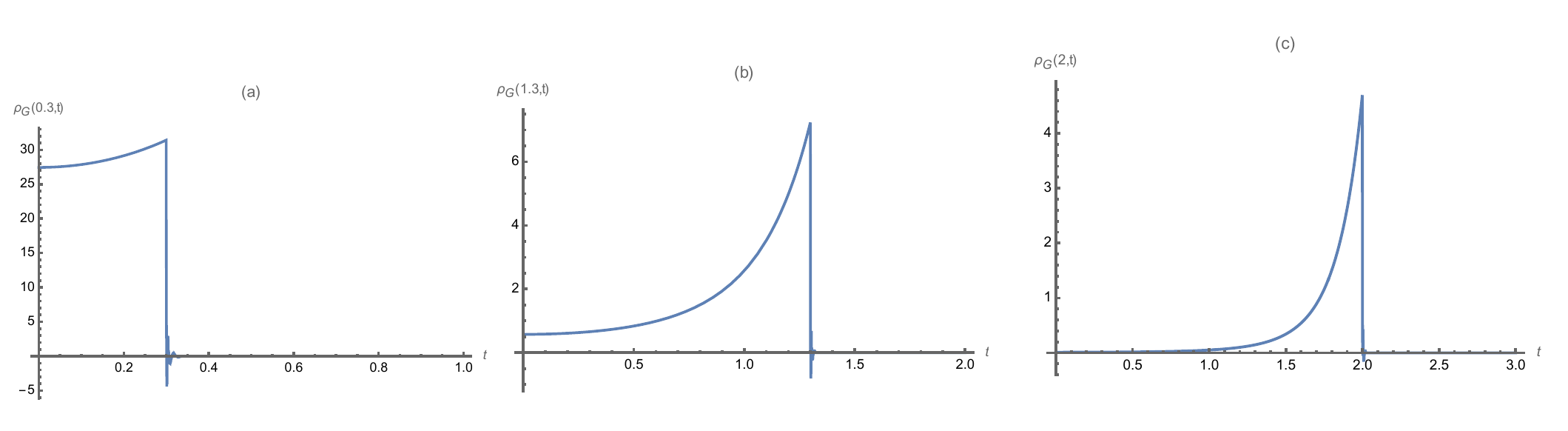}
\caption{Time dependent densities (a) $ \r_G(.3,t) $ (b) $\r_G(1.3,t) $ (c) $\r_G(2,t)$. Units of $r$ and $t$ are relative to $R$.} 
\label{rt}\end{figure}
\end{widetext}

 %
 Perhaps this seeming contradiction could be resolved by  considering   time dependence. Consider the matrix element of the time component of the current as a function of the four-position $x=(t,\bfr),\,j^0(x)$, in the wave packet used by  Ref.~\cite{PhysRevLett.129.012001}. Then the space-time density is given by
\begin{widetext}\bea
\r(x)& 
&=\int {d^3p\over 2 E(p) (2\pi)^3}\int {d^3p'\over 2 E(p') (2\pi)^3}\frac{\la p'|j^0(0)|p\ra }{E(p)+E(p')} \phi(p) \phi(p') e^{i(p'-p)\cdot x}\,.
\eea \end{widetext}
 This is the same as Eq.(6) of Ref.~\cite{PhysRevLett.129.012001} except for keeping the  time dependence. Next, I follow 
 Ref.~\cite{PhysRevLett.129.012001}, and take their limit of the vanishing spatial extent of the wave packet. This leads to
 the result:
 \begin{widetext}\bea\label{tdep0}
 \r(t,r)=\int \frac{d^3q}{(2\pi)^3}e^{-i\bfq\cdot\bfr}\int_{-1}^1d\a{1\over 2}F((\a^2-1)\bfq^2)e^{i \a |\bfq| \,t}
 ,\eea\end{widetext}
 which is the same as Eq(11) of Ref.~\cite{PhysRevLett.129.012001} in the limit that $t=0$. The only difference with that reference arises from appearance of  the exponential $e^{i \a |\bfq| \,t}$ that appears as a result of the same logic used to obtain the $\a^2$ term of \eq{tdep0}.

 Let's consider the limit of very large times. In that case, the integrand oscillates rapidly and the integral vanishes. Thus one expects  that 
  \bea 
  \lim_{t\to\infty}\r(t,r)=0.
  \eea
 The density seems to vanishe as time increases. This surely represents a problem with using \eq{ulf1} to compute form factors.
  The result is a vanishing density for $r\ne t$. 

 Further understanding may be obtained by examining   \eq{tdep0} using an asymptotic expansion in large, positive values of $t$. This is done by using
 $e^{i\a |\bfq|\,t}= {1\over i |\bfq|\,t}{\partial e^{i\a |\bfq|\,t}\over \partial\a}$ and integrating  over $\a$ by parts. Keeping only  the leading  term in powers of $1/t$
 leads to 
 \bea \r(t,r)\sim {1\over 4\pi^2 r}\int \frac{d^3q}{(2\pi)^3}{1\over  |\bfq|\,t}e^{-i\bfq\cdot\bfr}\sin{  |\bfq|\,t}
 \eea
 because $F(0)=1$.  The integration over angles of $\bfq$ yields
  \bea &\r(t,r)\sim {1\over \pi r\,t}\int {dq} \sin|\bfq| \,r  \sin{  |\bfq|\,t}\\&
  =   \frac{1}{8\pi\, rt}  \d (r -t).
  \eea
At large times, $t$, the density vanishes except at $r=t$.

It is useful to study the full time-dependence by using the  specific example of a
   Gaussian form factor, 
 \bea F_G((\a^2-1)\bfq^2)= e^{-(1-\a^2)\bfq^2 R^2/6}.\eea
 A closed form of the expression for the resulting density at $t=0$ is obtained by doing the integral of \eq{tdep0} to find
\bea
\r_G(0,r)= \frac{3 e^{-\frac{3 r^2}{2 R^2}}}{2 \pi  r R^2}.
\eea
The $1/r$ behavior at the origin seems to be a remarkably unphysical way to account for a simple Gaussian form factor.

 Next  let's examine the time dependence of $\r_G(t,r)$. The integral over $q$ in \eq{tdep0} may be done in closed form with the result
\bea& \r_G(t,r)={3\over 4r R^3} \sqrt{\frac{3}{2\pi^3}}\nonumber\\&
\times
\int_{-1}^1 {d\a \over \left(1 -  \a^2\right)^{3/2}} (F(r-\a t)+F(r+\a t)),\\&
F(x)\equiv x e^{-3 x^2\over \sqrt{1-\a^2}}
   .  \eea
Proceed by numerical integration to find $\r_G(r,t)$ with  $r$ and $t$ in units of $R$. The results   in Fig.~\ref{rt} show that indeed the density vanishes for $t>r$.
Note the sharp falloff when $t=r$. The decreasing width as the value of $r$ increases indicates that the limit of a delta function is approached..  

The weird behavior at $r=0$ and the rapid fall off with time indicate that \eq{ulf1}
is simply not an expression that may be correctly used to compute a form factor of a system with relativistic dynamics. 

\subsection{Relativistic calculations of form factors}
 I therefore briefly pursue the  computation of  form factors for systems with relativistic dynamics, by considering the 
example, of a spinless system  made of two spin-0 constituents, the form factor is given by 
  \cite{Carbonell:2008tz}:
  \begin{widetext}
\bea (K+K')^\mu F(Q^2)=i\int {d^4k\over (2\pi)^4}(K+K'+k)^\mu(k^2-m^2)
\Phi({1\over2}K-k;K)\Phi({1\over2}K'-k;K'),\label{feynf}\eea
\end{widetext}
where $K'=K+q$. An example of an expression for $\Phi$ is given in \eq{ex}. The
only way in which \eq{feynf} can be massaged into the form of \eq{ulf1}  is in the  non-relativistic limit defined in the introduction~\cite{Miller:2009fc,Miller:2009sg}.
An  example of the  disparate forms, note that one obtains instant-form expressions by integrating over $k^0$. There are poles for both positive and negative values of $k^0$. Poles at negative energy cannot appear in \eq{ulf1}.

Once again, I find that there is no time-independent spherically-symmetric density for confined systems of relativistically moving constituents.
  \section{Summary}
  A method for obtaining time-independent two-dimensional densities is described in Sect.~II. The axial vector current and 
  mass density related to the trace of the energy momentum tensor (defining a mass density) are used as examples.
  
   That using the Breit frame violates the quantum mechanical definition of probability, the uncertainty principle and Poincar\'{e} invariance  is shown in Sect.~III. 
   
   The use of the Abel transformation to relate two and three dimensional distributions depends on a density having three-dimensional spherical symmetry. But densities of light-hadrons cannot be spherically symmetric; see Sect.~IV. 
   
    Wigner distributions provide wave packets with both vanishing average  size and average momentum. But the fluctuations in both position and momentum are infinite, as shown in Sect.~V. 
    
    The use of spherical wave packets of vanishing spatial extent leads to  densities, $\r(r,t)$,  that  vanish almost everywhere in space as time increases from an initial value; see Sec.~VI.
  
  The net result is that the use of the infinite momentum frame, light front formalism  seems to be the only way to obtain densities consistent with the quantum mechanical definition of probability, the uncertainty principle and the principle of relativity. 
   \section*{Data availability}
   No data was produced or used.
  \section*{Acknowledgements}
 I  gratefully acknowledge the hospitality of  MIT LNS, UC Berkeley Physics Department, Lawrence Berkeley Lab and Argonne National Lab while this work was done. 
 
  I thank Adam Freese for valuable comments and for his collaboration on several related papers. I thank Keh-Fei Lui, Zein-Eddine Meziani and Dimitra Pefkou for useful discussions. I thank Phiala Shanahan and Dimitra Pefkou for providing the gravitational form factors used in Sect.~II. The research in this work received inspiration from the goals of the Quark Gluon Tomography Topical Collaboration of the
U.S. Department of Energy. 
  \maketitle     
\noindent

 \input{DownWithBreitFrame.bbl}
 
\end{document}

%% file: DownWithBreitFrame.bbl
%